\documentclass{appolb}
\usepackage{epsfig}

\begin{document}
\title{POSSIBLE SIGNAL FOR CRITICAL POINT IN HADRONIZATION PROCESS}
\author{
M.Rybczy\'nski, Z.W\l odarczyk
\address{Institute of Physics, \'Swi\c{e}tokrzyska Academy,\\
 \'Swi\c{e}tokrzyska 15; 25-406 Kielce, Poland;\\
 e-mail: mryb@pu.kielce.pl and
wlod@pu.kielce.pl}
\and  G.Wilk
\address{The Andrzej So\l tan Institute of Nuclear Studies,\\
   Ho\.za 69; 00-689 Warsaw, Poland; e-mail: wilk@fuw.edu.pl}
}
\maketitle

\begin{abstract}
We argue that recent data on fluctuations observed in heavy ion
collisions show non-monotonic behaviour as function of number of
participants (or "wounded nucleons") $N_W$. When interpreted in
thermodynamical approach this result can be associated with a
possible structure occuring in the corresponding equation of state
(EoS). This in turn could be further interpreted as due to the
occurence of some characteristic points (like {\it softest point} or
{\it critical point}) of EoS discussed in the literature and
therefore be regarded as a possible signal of the QGP formation in
such collisions. We show, however, that the actual situation is still
far from being clear and calls for more investigations of fluctuation
phenomena in multiparticle production processes to be performed. 
\end{abstract}
\PACS{24.60.Ky; 25.75.Nq; 05.20.-y}

\section{Introduction}

The primary goal of all high energy heavy ion experiments is to find
the phase transition between hadronic matter and quark-gluon plasma
(QGP) as predicted by the lattice QCD \footnote{For relevant
references see, for example, proceedings on recent QM conference
\cite{QM}.}. It is natural then to look at experimental results from
such perspective  and to scrutinize them for the possible signals of
QGP formation. In this  work we would like to concentrate on one
characteristic feature, which  could signal such phase transition,
namely it should be accompanied by  enhanced fluctuations in some
variables \cite{GGM,SSS,THM,LML}. For it turns out that, when
measuring in final states only hadronic probes produced with small 
transverse momenta, the only observables surviving the possible phase
transition (and, in principle, actually depending on its form) is the
observed fluctuation pattern in properly selected variables.
Following \cite{GGM} we shall be interested in the quantity being the
ratio of fluctuations in entropy to fluctuations in energy. The
reason is that, in thermodynamical approach to be pursued here, both
quantities are well defined for any form of matter (confined, mixed
and deconfined) both in early stage of collisions and in the final
state of the hadronizing system under consideration. When produced
matter can be treated as an isolated system, energy is conserved and
entropy is also expected to be conserved during the expansion and
freezeout stages of the hadronization. Therefore simultaneous
measurements of both quantities are likely to provide us with
information on the equation of state (EoS) of the hadronizing system.
This ratio can be written (see Appendix A for details) as
\begin{equation}
R = \frac{\left(\frac{\delta S}{S}\right)^2}
         {\left(\frac{\delta E}{E}\right)^2} = 
         \left( 1 + \frac{\alpha}{1 + \alpha \phi}\right)^{-2},\qquad
         {\rm where}\qquad \phi = \frac{d \ln V}{d \ln T} = 
         \frac{T}{V}\frac{dV}{dT}.         \label{eq:R}
\end{equation}
Here $\alpha = dp/d\varepsilon = p/\varepsilon$ characterizes EoS, $p$ 
denotes the pressure and $\varepsilon =E/V$ energy density with $V$ being 
the volume of our system and $T$ its temperature (statistical equilibrium 
is assumed). The entropy of the system, $S$, should be connected 
with the  multiplicity of the produced secondaries, $S \sim N$
and, similarly, energy released in the production process, $E$, 
should be connected with the measured sum of transverse momenta, 
$E \sim \sum_{i=1}^N p_T$. When variations of the reaction volume 
(identified with the volume of system) can be neglected, i.e., for 
$\phi = 0$, one gets formula proposed in \cite{GGM} as the right quantity
the energy dependence of which should be investigated experimentally
\footnote{For recent reviews on this matter see \cite{GG}.}:
\begin{equation}
R = \frac{\frac{\left(\delta S\right)^2}{S^2}}
      {\frac{\left(\delta E\right)^2}{E^2}}
= \left(1 + \frac{p}{\varepsilon}\right)^{-2} = 
\frac{1}{\left( 1 + \alpha\right)^2}. \label{eq:RF}
\end{equation}
Notice that $R$ is directly sensitive to the equation of state (EoS)
of hadronizing matter because of occurrence of parameter $\alpha$. 
In what follows we shall assume that 
\begin{equation}
\frac{\delta S}{S} =  \frac{\delta N}{N}\quad {\rm and}\quad
\frac{\delta E}{E} = \frac{\delta \sum p_T}{\langle \sum p_T\rangle}
. \label{eq:assume}
\end{equation} 
We identify throughout this work $\delta N$ and $\delta \sum p_T$
with fluctuations caused by {\it all possible sources} (as only such
are accessible in experiment), i.e., also all our results concerning
different estimations of $p/\varepsilon$ by means of eq.
(\ref{eq:RF}) include influences of all possible sources of fluctuations.
In \cite{GGM} $R$ has been investigated by using some specific
statistical model of nuclear collisions. The predicted very
characteristic shape of energy dependence of this quantity
(non-monotonic, with a single "shark fin"-like maximum) was then
proposed as a possible signal of QGP phase transition which should
therefore be subjected to future experimental tests \cite{GG}.

In this work we would like to bring attention to the fact that
apparently similar kind of fluctuations (but in function of $N_W$,
which can be connected with the reaction volume, rather than energy,
as is the case in \cite{GGM}) have been already observed and reported
in \cite{PHENIX} and \cite{NA49}. In what follows, using some minimal
input, we shall in the next Section rewrite results of
\cite{PHENIX,NA49} in terms of variable $R$ defined by (\ref{eq:RF})
and (\ref{eq:assume}), clearly demonstrating that it has similar 
non-monotonic character as that obtained in \cite{GGM}. It is
therefore tempting to argue that these data show us either what has
been called in the literature the {\it softest point} of the
corresponding EoS \cite{HS} or what is named in other investigations
the {\it critical point} of EoS \cite{LAT}.  However, at this stage
one not only cannot distinguish between both  possibilities but even
one cannot take them very seriously, for when confronting these data
with similar data obtained recently in \cite{CERES},  which use
different measure of fluctuations, one discoveres that  no such
feature are seen there\footnote{Ref. \cite{CERES} provides  data for
parameter $\Phi$ (defined in \cite{PHI}) at energies $40$,  $80$ and
$158$ AGeV for the following centrality bins: $0-5$\%,  $0-6.5$\%,
$5-10$\%, $10-15$\% and $15-20$\%.}. Still, there are some  other
possibilities offered recently, as, for example, spinodial
decomposition occuring at the hadronization stage \cite{Spinod},
which  deserve also attention. 

In Section III we shall also analyse (with the same aim as above, i.e., 
looking for a possible signal of QGP) recent interferometric data,
which provide us information on the interaction volume
\cite{PRL} \footnote{The volume $V$ measured here is
the freeze-out volume but in our discussion we identify
it with the production volume.}.
It turns out that they lead to similar results for
EoS as data from \cite{PHENIX,NA49} (albeit this time at different energy). 
The last Section contains our summary whereas Appendices contain
derivation of our basic formulas.

\section{Nonmonotonical dependences observed in data on fluctuations} 

In \cite{PHENIX} the following measure of fluctuations has been
presented as a function of the number of participants $N_W$ (or "wounded"
nucleons)\footnote{For proper definition of different types of averages
and mean values used here see \cite{FOOTB}.}:  
\begin{equation}
F_{T} = \frac{\Omega_{data} -
\Omega_{random}}{\Omega_{random}};\quad \Omega=\frac{\sqrt{\langle
\bar{p}^2_T\rangle - \langle \bar{p}_T\rangle^2}}{\langle \bar{p}_T\rangle}
.\label{eq:FT} 
\end{equation}
It increases with $N_W$, reaches maximum at $N_W\sim 200$, and
decreases for higher values of $N_W$. Whereas in \cite{PERCOL} such
behaviour has been attributed to the peculiar feature of the
hadronization model used, namely to the percolation of hadronizing
strings produced in collision process\footnote{Without introducing 
notion of EoS or phase transitions, unless phenomenon of percolation 
itself is regarded as a kind of phase transition. But even
then it would be phase transition between bigger
and smaller number of strings only, not between
hadronic matter and quark-gluon plasma.}, here we shall
connect it directly, in a similar way as in \cite{GGM}, to the
behaviour of EoS of the matter produced at the early stage of the
collision.
This can be done by rewritting $F_T$ in terms of $R$ defined in eq.
(\ref{eq:RF}). To do so let us first notice that, because  
\begin{equation}
\frac{Var\left(\sum p_T\right)}{\langle \sum p_T\rangle^2} = 
\frac{Var\left( p_T\right)}{\langle p_T\rangle^2}\frac{1}{\langle
N\rangle} + \frac{Var(N)}{\langle N\rangle^2} , \label{eq:a}
\end{equation}
one can write $R$ as
\begin{equation}
R = \frac{\frac{Var(N)}{\langle N\rangle^2}}
     {\frac{Var\left(\sum p_T\right)}{\left\langle \sum p_T\right\rangle^2}}
    = \frac{1}{1 + \frac{1}{r}}, \label{eq:b}
\end{equation}
where
\begin{equation}
r = \frac{1}{\omega}  \frac{Var(N)}{\langle N\rangle}\quad {\rm and}\quad 
      \omega = \frac{Var(p_T)}{\langle p_T\rangle^2}. \label{eq:NpT}
\end{equation}
On the other hand, as shown in Appendix B, $F_T$ can be expressed in
terms of $r$ and parameter $\rho$,
\begin{equation}
F_T = \sqrt{1 + 2r - 2\rho \sqrt{r^2 + r}}-1 , \label{eq:FRF}
\end{equation}
where $\rho \in[-1,1]$, the correlation coefficient between number of
particles $N$ and $\sum p_T = \sum_{i=1}^{N} p_T$, is our minimal
input mentioned above\footnote{It appears because, whereas $R$
depends on fluctuations in $N$ {\it and} on fluctuations in
$\sum_{i=1}^{N}p_T$, so far in experiment one measures quantities
like $\Phi$ or $F_T$ in which both type of fluctuations occur
together. Parameter $\rho$ describes therefore their mutual (unknown
{\it a priori}) correlations.}. Using (\ref{eq:b}) it is now
straightforward to rewrite eq. (\ref{eq:FRF}) in the following form:  
\begin{equation}
\frac{(1+F_T)^2-1}{2} =
\frac{R}{1-R}\left[1-\frac{\rho}{\sqrt{R}}\right], \label{eq:fineq}
\end{equation}
from which our main result follows:
\begin{equation}
R = \frac{\rho^2 + 2y(1+y) \pm \rho\cdot\sqrt{\rho^2 + 4y(1+y)}}
                            {2(1 + y)^2}, \label{eq:Sol}
\end{equation}
where
\begin{equation}
y = \frac{1}{2}\left[ (1+F_T)^2 - 1\right] . \label{eq:pomoc}
\end{equation}
In addition to the {\it a priori} unknown correlation coefficient
$\rho$ there is also some freedom in the choice of sign in
(\ref{eq:Sol}). To fix both of them let us notice the
following:
\begin{itemize}
\item[$(a)$] It is known that purely statistical or broader
fluctuations encountered in all multiparticle production processes, 
i.e., the fact that $Var(N) \ge \langle N\rangle $, lead to the 
condition that\footnote{See, for example, \cite{Broad}. 
Multiplicity distributions broader than poissonian distributions 
are also observed in hadronic collisions \cite{NB}.}
\begin{equation}
\frac{Var(N)}{\langle N\rangle} = \frac{\omega}{\frac{1}{R}-1} \ge
1 .\label{eq:Poisbroad}
\end{equation}
\item[$(b)$] Furthermore, analysis of recent CERES data \cite{CERES}
shows clearly, see Fig. \ref{fig:Fig4} below, that parameter $\omega$ 
is very slowly varying function of energy and multiplicity
(esentially $\omega \simeq 0.43$). This practical constancy of
fluctuations in $p_T$ means that fluctuations of energy $E$, which
are given by fluctuations of $\sum p_T$,  is not so much given by
fluctuations in $p_T$ but by fluctuations in multiplicity $N$, cf.,
eq.(\ref{eq:a}). 
\end{itemize}
These two observations mean therefore that for the
multiplicity distributions of the poissonian type and broader, $R$
is limited from below: 
\begin{equation}
R > \frac{1}{1+\omega} > 0.699  \label{eq:limitR}
\end{equation}
(for $\omega = 0.43$). Using this limit together with eq.
(\ref{eq:Sol}) one obtaines that also $\rho^2$ is limited from below,
namely
\begin{equation}
\rho^2 > \frac{(\omega y -1)^2}{\omega +1} . \label{eq:limit}
\end{equation}
Because, without any additional correlations between $x_i$ and $N$, 
the $\sum_{i=1}^Nx_i$ increases with $N$, the positive correlation
coefficient, $\rho >0$, seems to be the only natural choice. In
addition, results of PHENIX \cite{PHENIX} show that $F_T$ is very
small (actually not exceeding $F_T\sim 0.04$). Neglecting therefore
term with $y$ in (\ref{eq:limit}) one can estimate that $\rho > 0.83$
(in our calculations we have put it slightly above this limit using
value $\rho = 0.85$). It means that $R$ corresponds to a stronger
than poissonian fluctuations. To be consistent with limitations on
$R$ imposed by (\ref{eq:limitR}) we shall choose to the solution
with positive sign in (\ref{eq:Sol}) (the other one would lead to
unacceptable small values of $R$).

\begin{figure}[ht]
\noindent
  \begin{minipage}[ht]{65mm}
    \centerline{
        \epsfig{file=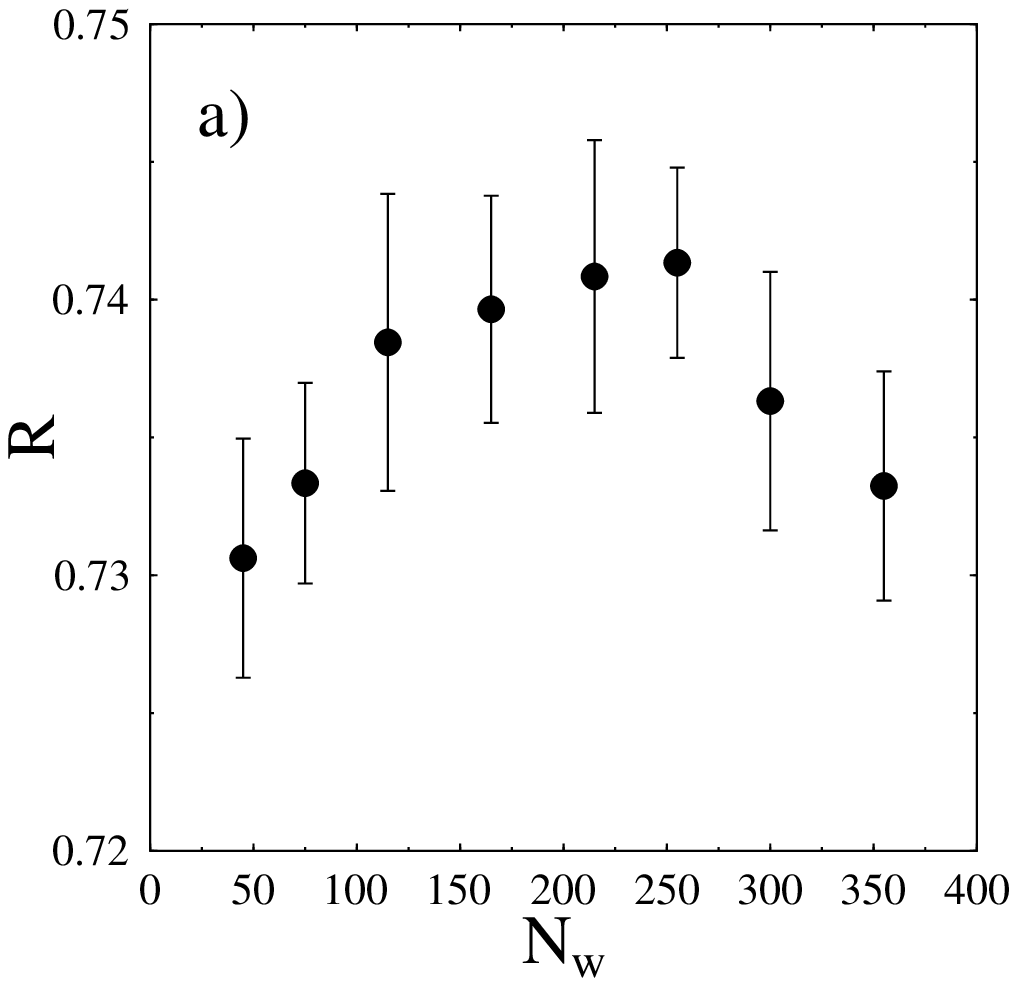, width=65mm}
     }
  \end{minipage}
\hfill
  \begin{minipage}[ht]{65mm}
    \centerline{
       \epsfig{file=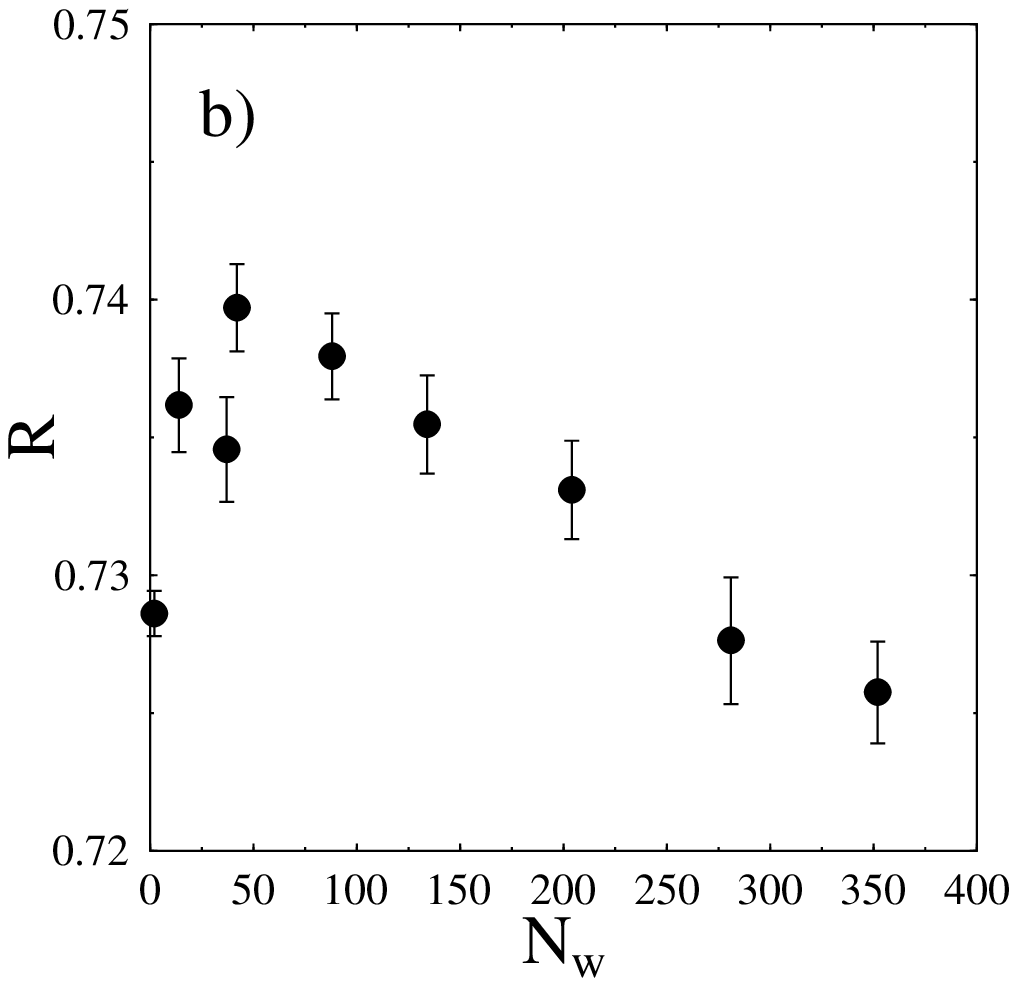, width=65mm}
     }
  \end{minipage}
  \caption{Left panel: results of transforming PHENIX data from
\protect\cite{PHENIX} (i.e., $F_T$ versus $N_W$) to $R$
as function of the number of participant (or number of "wounded")
nucleons, $N_W$ by using eq.(\ref{eq:Sol}). As explained in the text
the positive sign in (\ref{eq:Sol}) has been chosen and the
correlation coefficient set  to $\rho = 0.85$ (see text for
details). Right panel: $R$ obtained from $\Phi$ as measured 
by NA49 \protect\cite{NA49} (for the same value of $\rho$).} 
  \label{fig:Fig1}
\end{figure}

\begin{figure}[h]
 \begin{center}
  \begin{minipage}[ht]{90mm}
    \centerline{
        \epsfig{file=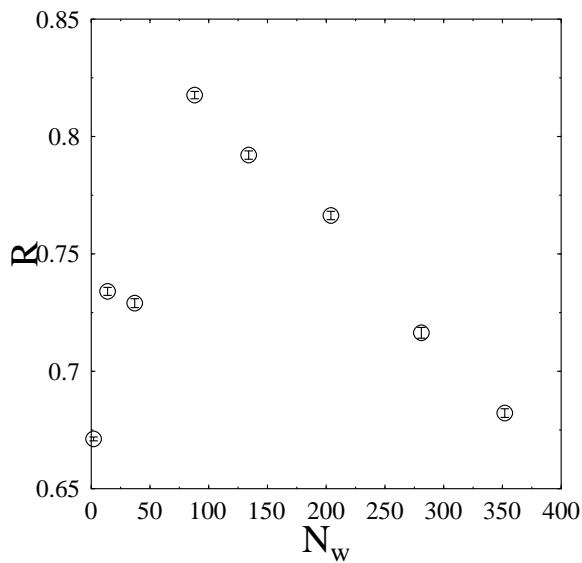, width=90mm}
     }
  \end{minipage}
 \end{center}
 \caption{The expected shape of $R$ as function of $N_W$ this
           time given directly by eq. (\ref{eq:b}) and data
           \cite{NA49}; no parameter $\rho$ enters here.
           } 
  \label{fig:Fig2}
\end{figure}

We are now prepared to translate, by using eq. (\ref{eq:Sol}), the
experimental knowledge of $F_T$ and $\Phi$ as a function of number of
participants provided in \cite{PHENIX,NA49} into similar dependence of
$R$. The results are presented in Fig. \ref{fig:Fig1}. The only unknown 
feature there is the choice of parameter $\rho$. In fact for $\rho=$const
different values result in essentially the same shape of $R$, only shifted
accordingly paralelly up or dow. The case of $\rho = \rho(N_W)$ would
lead to some changes, however. Unfortunately at the moment there are no 
data available to estimate the functional form of $\rho(N_W)$. We expect,
however, that our choice of $\rho$ used here corresponds to a 
lower limit for a possible effect observed in Fig. \ref{fig:Fig1}.
To substantiate this we present in Fig. \ref{fig:Fig2} parameter $R$
as function of $N_W$ calculated directly from definition
(\ref{eq:b}), i.e., using directly measured information on   
$Var(N)/\langle N\rangle$ and $Var(p_T)/\langle p_T\rangle^2$ obtained by
\cite{NA49}\footnote{This could be regarded as a possible suggestions
for experimentalists that everything needed for such discussion as
presented here is measurable directly, without resorting to
quantities like $F_T$ or $\Phi$, see also \cite{PhiUWW} for similar
conclusions obtained at different circumstances. However, care must
be excercised when following such approach because usually bins in
centrality are large and one can expect therefore contributions to
fluctuations coming from fluctuations in read-offs of the ZDC
calorimeter, which for a time being are not know.}.  

So far we have not yet used connection of $R$ with EoS variables
$p$ and $\varepsilon$ as given by eq. (\ref{eq:RF}). According to it
we can interpret the peculiar shape of curve obtained in Fig.
\ref{fig:Fig1} in terms of the type of EoS admitting it, in
particular as due to a specific behaviour of $p/\epsilon$. Following
therefore discussion in \cite{GGM} we argue that this shape could be
regarded as a signal of the existence of  
either the {\it softest point} of the corresponding EoS
\cite{HS} or its {\it critical point} \cite{LAT}.
This time it would show up as function of the number of participants
(which can be translated into a volume                                      
of the reaction) rather than energy, as discussed in \cite{GGM}.

\begin{figure}[ht]
\noindent
  \begin{minipage}[ht]{65mm}
    \centerline{
        \epsfig{file=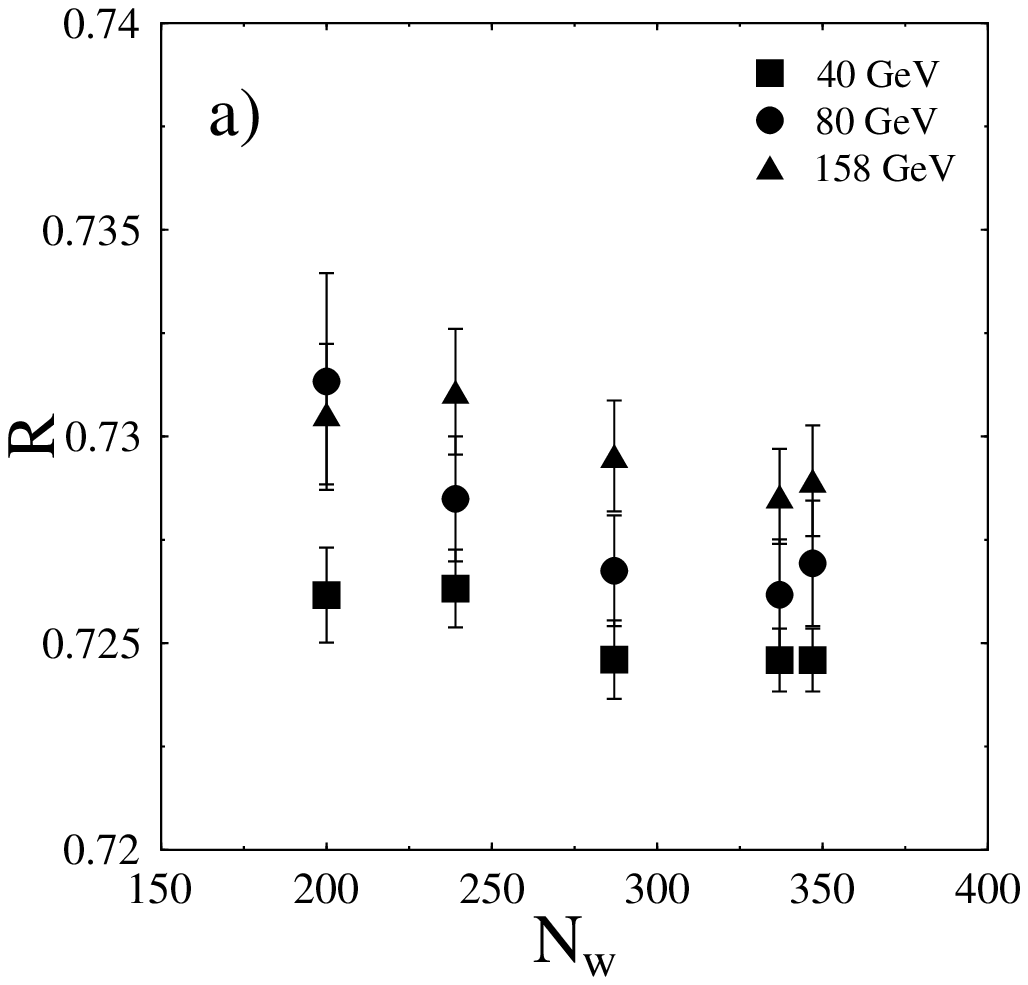, width=65mm}
     }
  \end{minipage}
\hfill
  \begin{minipage}[ht]{65mm}
    \centerline{
       \epsfig{file=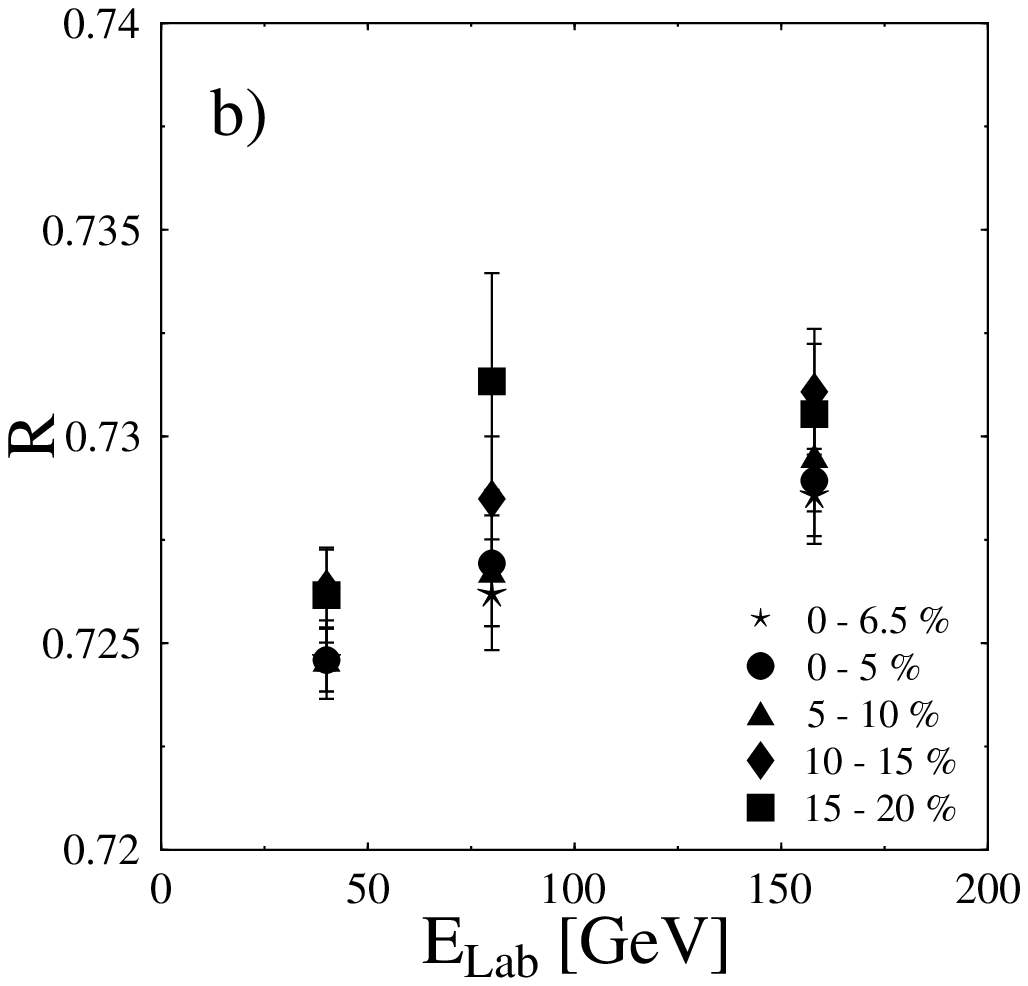, width=65mm}
     }
  \end{minipage}
  \caption{The same as in Fig. \protect\ref{fig:Fig1}b but this time
using recent data on $\Phi$ parameter obtained by CERES Collaboration
\protect\cite{CERES}. The results are presented both as function of
participants ($(a)$) and as function of energy ($(b)$).} 
  \label{fig:Fig3}
\end{figure}

Actually, because of relation between $F_T$ and measure $\Phi$ of 
fluctuations \cite{PHI} (see (\ref{eq:FTPHI}) in Appendix B), 
measurements of variable $\Phi$ is equally good for the kind 
of analysis performed here. This is clearly seen in Fig. 
\ref{fig:Fig1}b and \ref{fig:Fig2} using NA49 data \cite{NA49} and in
Fig. \ref{fig:Fig3} where recent data on $\Phi$ (cf., \cite{CERES}) 
were used as our input\footnote{Notice that in both cases the lack of
fluctuations in $N$, i.e., $Var(N) =0$ would immediately result in
$\Phi=F_T=0$. Lack of fluctuations means that also $\rho=0$ what
(together with $y=0$, cf. (\ref{eq:pomoc})) results in $R = 0$ as
well.}. In Fig. \ref{fig:Fig4} we show that data from \cite{CERES}
clearly indicate that, as was already mentioned before, fluctuations
of transverse momentum defined by variable $\omega$ introduced in eq.
(\ref{eq:NpT}) depend very weakly both on the energy and on the
centrality of the collision (i.e., on the number of struck nucleons
$N_W$). Therefore fluctuations of transverse momenta are practically
irrelevant for the problem considered here, i.e., for the apparent
structure seen in the EoS. Main effect is provided by fluctuations of
multiplicity $N$, as presented by eq. (\ref{eq:b}).

Let us close this Section with the following remark. Because
fluctuations in $\sum p_T$ are tightly connected with fluctuation of
temperature $T$, therefore one can write that $\frac{\delta E}{E} =
\frac{\delta T}{T}$ and express parameter $R$ in yet another form: 
\begin{equation}
R = \frac{\frac{Var(N)}{\langle N\rangle^2}}
                           {\frac{Var(T)}{T^2}} . \label{eq:flucT} 
\end{equation}
As shown in \cite{WWq} \footnote{For other hints on nonextensivity in
hadronic production processes and references to nonextensive 
statistics, see \cite{WWqq}.} relative fluctuations of temperature 
$T$ are connected with the heath capacity and can be parameterized 
by the nonextensivity parameter $q$,
\begin{equation}
\frac{Var(T)}{T^2} = \frac{N_W}{C_V} = q -1 ,\label{eq:CVq}
\end{equation}
which in our case is given by 
\begin{equation}
q-1=\frac{1}{\langle N\rangle}\left[ \omega + \frac{Var(N)}{\langle
N\rangle}\right] .\label{eq:qvar}
\end{equation}
\vspace{-5mm}
\begin{figure}[ht]
\noindent
  \begin{minipage}[ht]{60mm}
    \centerline{
        \epsfig{file=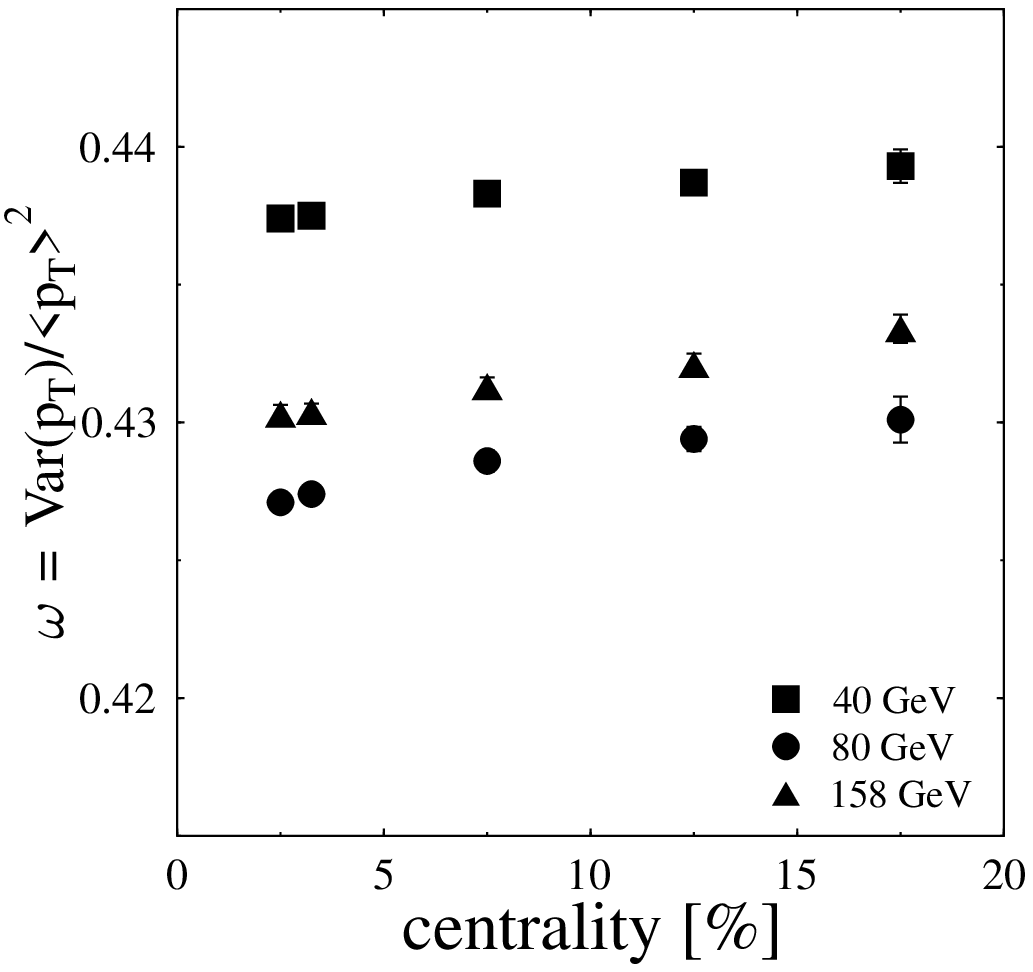, width=60mm}
     }
  \caption{Fluctuations of the transverse momentum as defined in eq.
(\ref{eq:NpT}) as function of centrality and for different
energies (data are from \protect\cite{CERES}, they are presented
before removal of short range correlations).} 
  \label{fig:Fig4}
  \end{minipage}
\hfill
  \begin{minipage}[ht]{60mm}
    \centerline{
       \epsfig{file=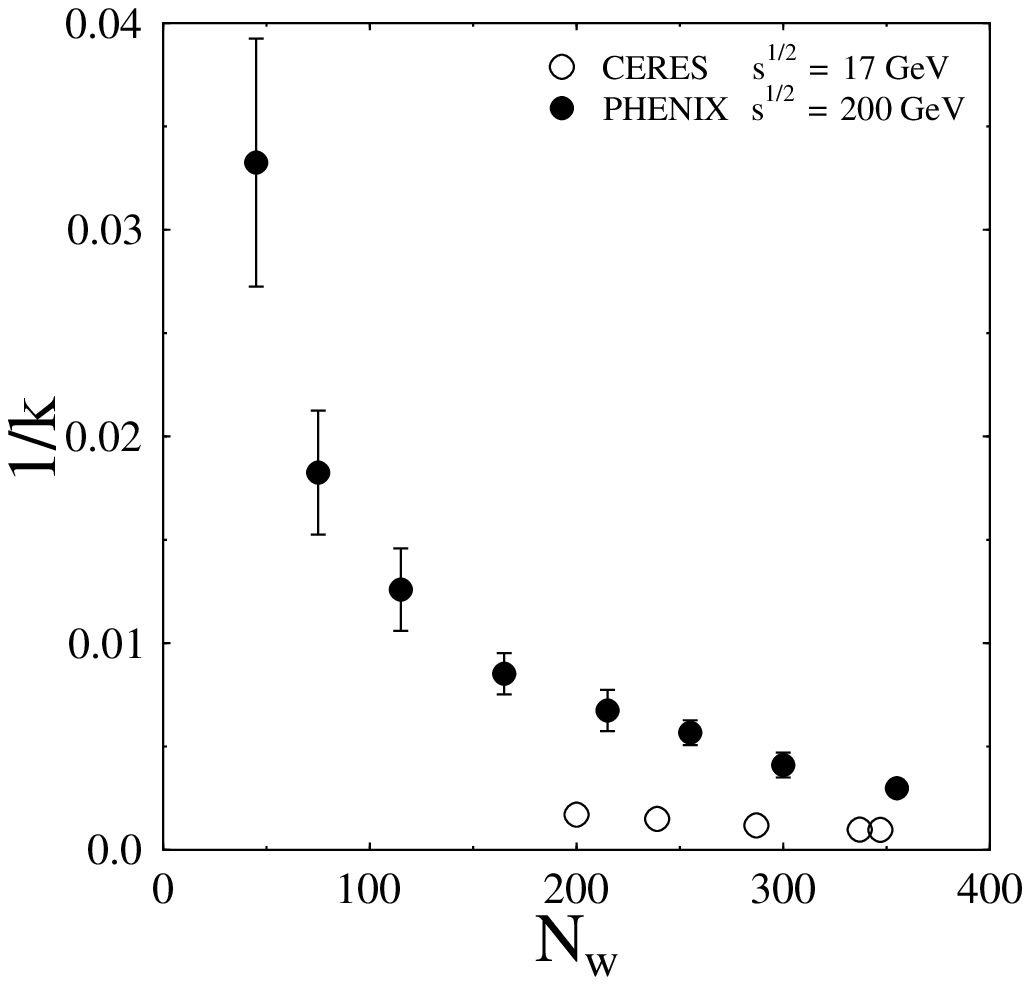, width=60mm}
     }
  \caption{Dependence of the NB \protect\cite{NB} multiplicity
distribution parameter $1/k$ on the number of participants $N_W$ for
PHENIX \protect\cite{PHENIX} and CERES \protect\cite{CERES} data.
\vspace{8mm}} 
  \label{fig:Fig5}
  \end{minipage}
\end{figure}

Notice that results shown in Fig. \ref{fig:Fig4}, i.e., that
essentially $\omega=const$, means that $q$ depends only on the
multiplicity (or on the number of struck nucleons), in fact $q-1$
diminishes as $1/\langle N\rangle$. Parameter $q$ allows us then (by
using eq. (\ref{eq:NpT})) to express parameter $R$ in the form
stressing its vital dependence on fluctuations of multiplicity, i.e.,
on $Var(N)/\langle N\rangle$:
\begin{equation}
R = \frac{1}{q-1}\frac{Var(N)}{\langle N\rangle^2} .
\end{equation}
As is well known these fluctuations lead to deviations from the
poissonian form of multiplicity distributions in multiparticle
production processes and result in broader distributions usually
expressed by the so called  Negative Binomial (NB) form and
characterized by the parameter $k$ \footnote{Cf. \cite{NB} for
details. Connection between NB and nonextensivity in hadronic
collisions represented by $q>1$ has been discussed in \cite{INELK}.},
which in our case is given by: 
\begin{equation}
\frac{1}{k} = \frac{Var(N)}{\langle N\rangle^2} - \frac{1}{\langle
N\rangle} = R(q-1) - \frac{1}{\langle N\rangle} . \label{eq:NB}
\end{equation}

Its dependence on $N_W$ for PHENIX data (at $\sqrt{s} = 200$ GeV) and
for CERES data (at $\sqrt{s} = 17$ GeV) is shown in Fig.
\ref{fig:Fig5}. As seen there parameter $1/k$ decreases with 
increasing number of participants. It means that (cf. \cite{NB}) with
increasing centrality fluctuations of the multiplicity become weaker
and the respective multiplicity distributions approach poissonian
form. Notice that because $R<1$ the following condition,
\begin{equation}
\frac{1}{k} + \frac{1}{\langle N\rangle} < q - 1 , \label{eq:condition}
\end{equation}
must be satisfied.

\section{Nonmonotonicity and phase transition}

\begin{figure}[ht]
\noindent
  \begin{minipage}[ht]{65mm}
    \centerline{
        \epsfig{file=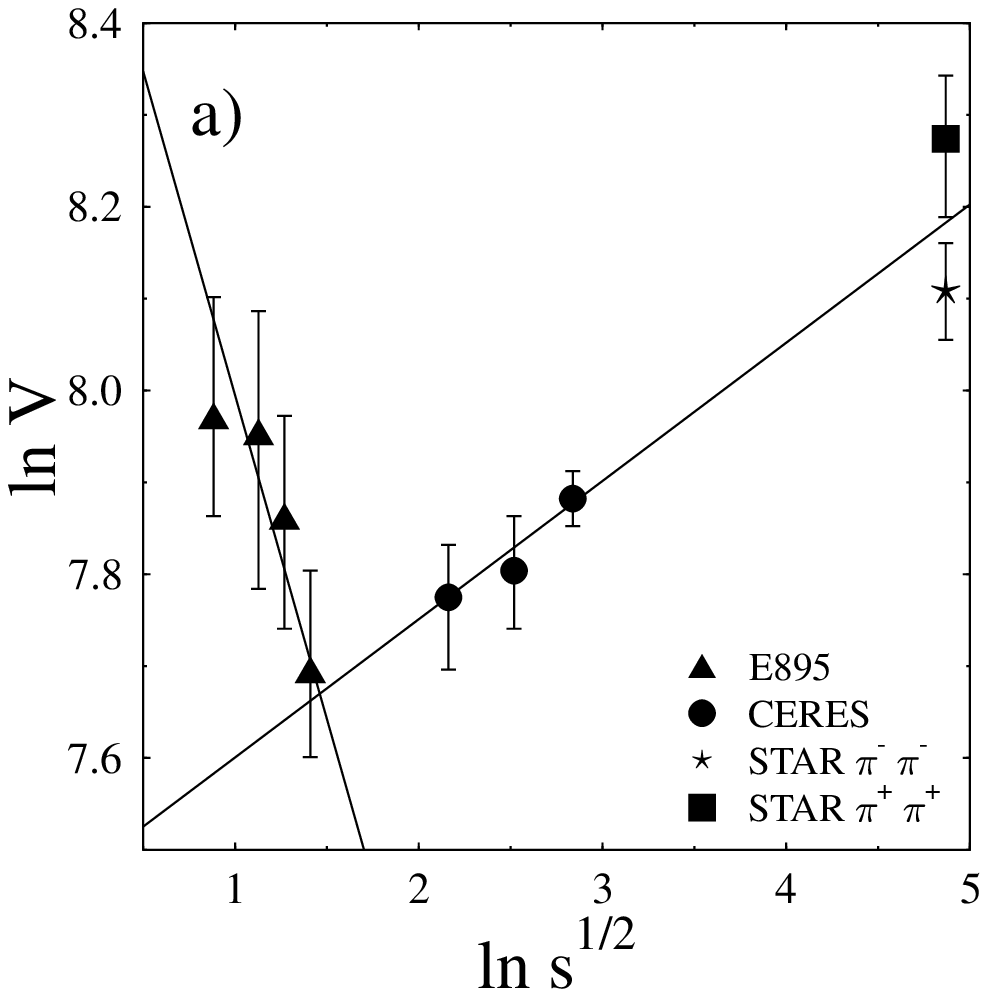, width=65mm}
     }
  \end{minipage}
\hfill
  \begin{minipage}[ht]{65mm}
    \centerline{
       \epsfig{file=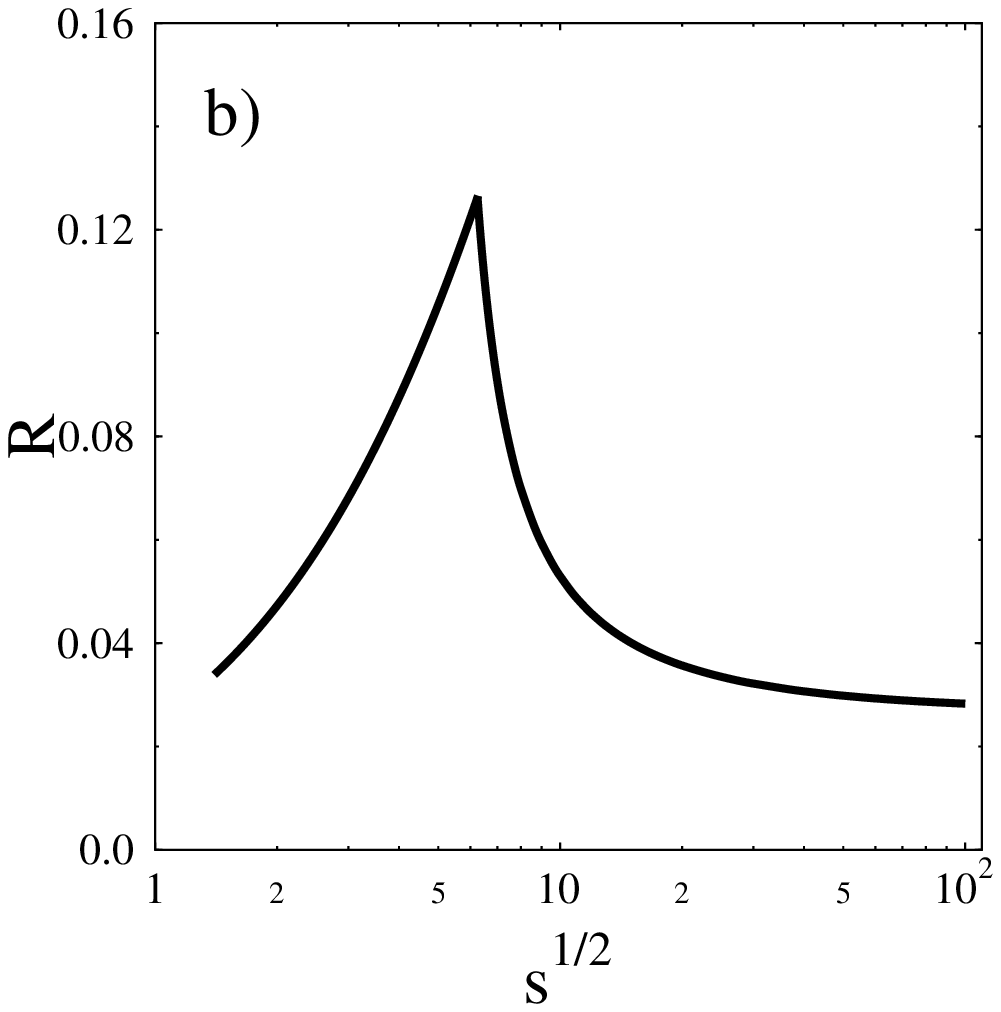, width=65mm}
     }
  \end{minipage}
  \caption{$(a)$ Our fits to data on volume of the interaction region
$V$ for different energies \protect\cite{PRL}; $(b)$ the
corresponding $R(\sqrt{s})$ dependence.} 
  \label{fig:Fig6}
\end{figure}

Let us now continue discussion of the parameter $R$ from the point of
view of its possible connection with the phase transition. Using
known thermodynamical identities \cite{Huang} one can rewrite $R$
(see Appendix C, all derivatives are for $T=const$ and $p = \alpha
\varepsilon$) as
\begin{equation}
R = \frac{\left(\frac{\partial \ln V}{\partial \ln E}\right)^2}
    {1 - C_VT\frac{\alpha}{E}\frac{\partial \ln V}{\partial \ln E}} .
\label{eq:RRRR}
\end{equation}
Notice that because in the vicinity of critical point
$\left(\partial p/\partial V\right)_T \rightarrow 0$ the parameter
$R$ reaches its maximum there:
\begin{equation}
R \rightarrow \frac{\varepsilon^2}
                   {\left(\frac{\partial E}{\partial
V}\right)^2_T} = \left(\frac{\partial \ln V}{\partial \ln E}\right)^2_T
.\label{eq:limite} 
\end{equation}
Let us discuss eq. (\ref{eq:RRRR}) in more detail concentrating on
two cases important for us. First of all notice that for fixed energy
$\sqrt{s}$ we have $\xi dE/E = dV/V = dN_W/N_W$ and with increasing
$N_W$ the value of $R$ decreases for $\xi =1$ and increases for $\xi
= -1$. It means then that for fixed $N_W$ parameter $R$ goes through
its maximum when $\partial \ln V/\partial \ln E$ changes sign.
Actually exactly such change of sign is observed in compilation of
heavy ion data on central collisions (for which $N_W$ remains
approximately fixed) \cite{PRL}, see Fig. \ref{fig:Fig6}a, from which
one can deduce that  

\begin{eqnarray}
\frac{\partial \ln V}{\partial \ln E} = \left\{\begin{array}{lll}
c_1 = -0.73 & \textrm{for} & s < s_c \\
c_2 = +0.15 & \textrm{for} & s> s_c
\end{array} \right . \label{eq:results}
\end{eqnarray}

Using now these values together with eq. (\ref{eq:RRRR}) we have
obtained dependence of $R$ on energy $\sqrt{s}$, cf. Fig.
\ref{fig:Fig6}b. The parameters used were such that
$$\frac{N_W\alpha T}{q-1}\frac{2}{N_W} = \frac{2\alpha T}{q-1} = 30$$
what corresponds to the reasonable values of $\alpha =1/3$, $T=0.14$
GeV and $q-1\approx 1/k = 0.003$ \footnote{One should notice that in
eq. (\ref{eq:RRRR}) we have total energy: $E=\sqrt{s}N_W/2$. This is
the origin of appearance of $N_W$ in presenting $R$ as function
of $\sqrt{s}$.}. Notice that $R$ shown there reaches its maximum for
energy $\sqrt{s_c} \simeq 6$ GeV. This "critical" value of energy  is
given by 
\begin{equation}
\sqrt{s_c} = \frac{\alpha T}{q-1}\frac{2}{N_W}\frac{c_1c_2}{c_1 +
c_2} , \label{eq:sc}
\end{equation}
where $c_i$ denote the corresponding values of derivatives $\partial
\ln V/\partial \ln E$ as given in (\ref{eq:results}).
Actually, although the shape of $R$ in Fig. \ref{fig:Fig6}b
resembles that expected in \cite{GGM}, the height of its maximum is
much smaller than expected there (to get the value observed in
\cite{GGM} one would have to shift it upward by $\sim 0.7$). The only
possible explanation we could offer at the moment is the observation
that $R$ displayed in Fig. \ref{fig:Fig6}b does not contain
contribution from statistical fluctuations of multiplicity (which
would give precisely the seek for value of $0.69$ for the poissonian
fluctuations). It would then mean that what we are calculating here
are only changes in $V$. For $V=$ const we are therefore getting
$R=0$. On top of that one has poissonian fluctuations, which for
$V=$const provide, as mentioned above, the lacking $\sim 0.7$ on top
of what is observed. In obtaining it we have increased
fluctuations of multiplicity $N$ by poissonian component (adding to
$Var(N)$ the value of $\langle N\rangle$ for constant volume $V$), 
i.e., we have used here $R'$ instead of $R$ as our input information,
where $1/(1-R') = 1/\omega + 1/(1-R)$. It is worth to notice that
recent lattice calculations show behaviour of $R$ similar to obtained
by us here \cite{LAT}.  

\section{Summary and conclusions}

We have shown that the recently proposed method of (almost) direct
investigations of the EoS, especially finding a possible traces of
phase transitions to QGP phase of matter \cite{GGM} by analysing
energy dependence of some specific fluctuations observed in the
multiparticle production data obtained in heavy ion collisions, can  
be further extended to include also fluctuations of different types
than those proposed in \cite{GGM}. Three examples were discussed here:
$(i)$ recent PHENIX Collaboration data \cite{PHENIX} on $F_T$; $(ii)$
recent NA49 Collaboration data \cite{NA49} on $\Phi$ and $(iii)$ 
CERES Collaboration data \cite{PRL,CERES} on $\Phi$. In particular,
we have derived formula (our eq. (\ref{eq:Sol})) expressing parameter
$R$ introduced in \cite{GGM} by the measured quantity $F_T$ (defined
in (\ref{eq:FT}) and measured in \cite{PHENIX}). Our results (see
Fig. \ref{fig:Fig1}a) show that one observes similar characteristic
structure in $R$ as that expected in \cite{GGM} but now being present
at given fixed energy and for different centralities (expressed by
the number of participants $N_W$). As is seen in Fig. \ref{fig:Fig3}
and Fig. \ref{fig:Fig1}b, the same type of analysis can be performed
using as input the so called $\Phi$ measure of fluctuations as
provided by recent NA49 \cite{NA49} and CERES data \cite{CERES}. We
would like to stress here that because, as is witnessed by results
shown in Fig. \ref{fig:Fig4}, fluctuations in $p_T$ are rather
irrelevant here, our results are almost entirely due to the
fluctuations in multiplicity. It means therefore that they are not
sensitive to flow phenomena. 
        
We have also established connection of $R$ with fluctuations of
temperature $T$ described by nonextensivity parameter $q$ \cite{WWq}.
The fact that data \cite{PRL} clearly indicate that in the measured
centrality range fluctuations in transverse momentum are essentially
constant, see Fig. \ref{fig:Fig4}, allows us to express parameter
$R$ by combination of fluctuations of $T$ (i.e., by parameter
$q$) and fluctuations of multiplicity as given by the parameter $1/k$
of NB distribution \cite{NB}, see eq. (\ref{eq:NB}) and Fig.
\ref{fig:Fig5}.  

\begin{figure}[ht]
\noindent
  \begin{minipage}[ht]{65mm}
    \centerline{
        \epsfig{file=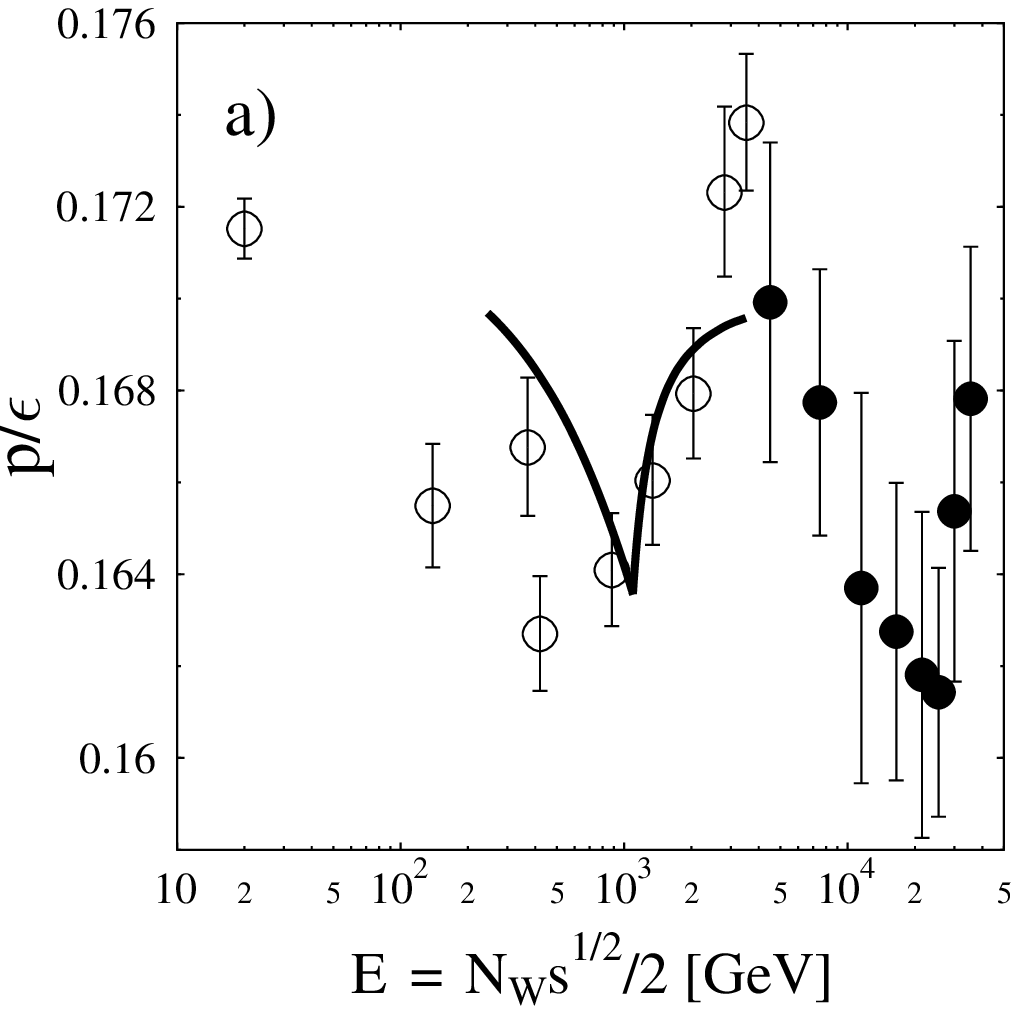, width=65mm}
     }
  \end{minipage}
\hfill
  \begin{minipage}[ht]{65mm}
    \centerline{
       \epsfig{file=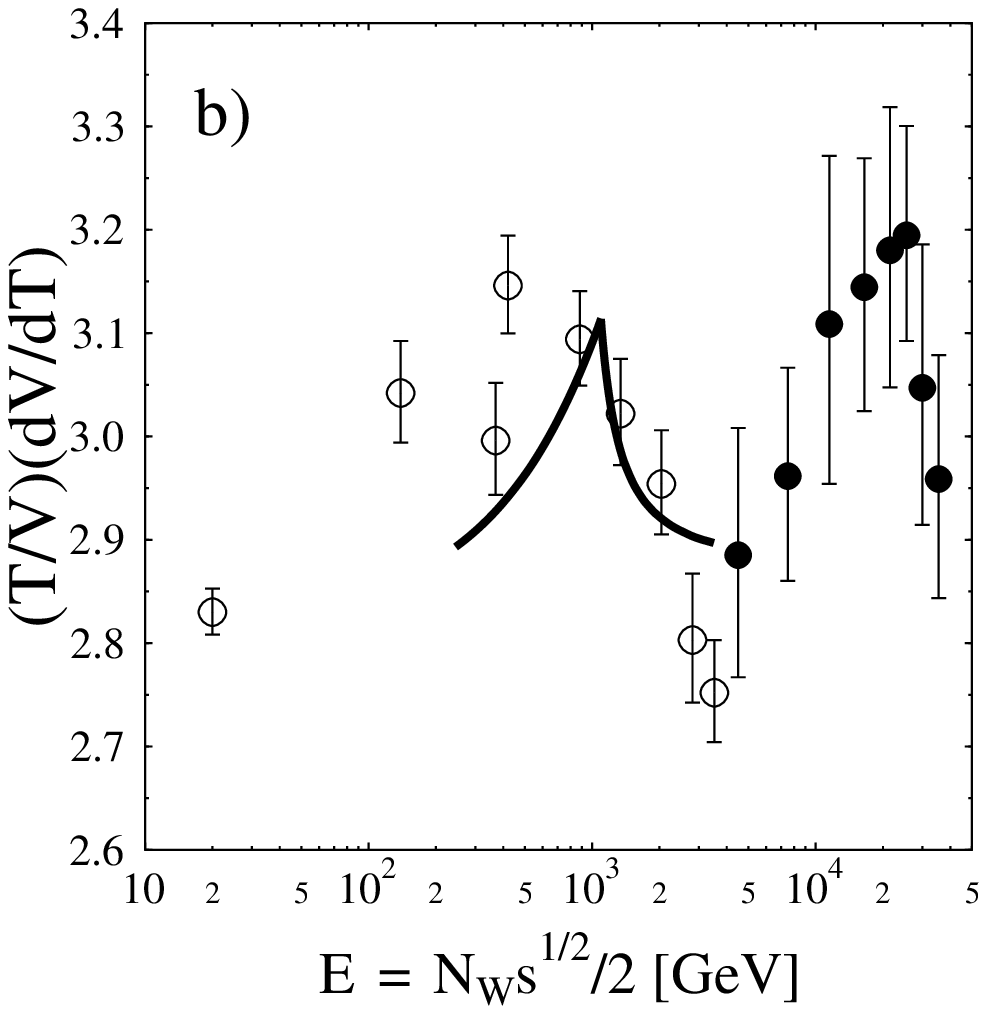, width=65mm}
     }
  \end{minipage}
  \caption{EoS corresponding to results obtained in Fig.
\ref{fig:Fig6}b (solid line) and to those obtained from Fig.
\ref{fig:Fig1} (points). Open points correspond to data from NA49
\cite{NA49} and full points to data from PHENIX \cite{PHENIX}. See
text for details. 
 }   
  \label{fig:Fig7}
\end{figure}

Finally, using some thermodynamical identities, we have expressed
$R$ in terms of parameter $\alpha = p/\varepsilon$ defining form
of EoS used and energy dependence of the reaction volume,
cf. eq. (\ref{eq:RRRR}). Using recent data on the later \cite{PRL}
(see Fig. \ref{fig:Fig6}a) we have obtained the characteristic
"shark fin" structure of the energy dependence of the parameter $R$
shown in Fig. \ref{fig:Fig6}b. It corresponds to the shape of EoS as
given in Fig. \ref{fig:Fig7}. Such structure of $R$
was predicted at \cite{GGM} (albeit for $\sqrt{s}$ replaced by the so
called Fermi collision energy measure) as reaching value of $R\sim
0.8$ at its maximum. In our case the maximum is much lower. Actually,
what is shown in Fig. \ref{fig:Fig7} are two situations, which can
emerge from definition of $R$ as given in eq. (\ref{eq:R}). Namely,
as one can notice variation of $R$ can originate either in variations
of $p/\varepsilon$ or in variations $dV/dT$ (in extremal cases). For
$dV/dT=0$ we are then getting from data $p/\varepsilon$, and this is
the case of Fig. \ref{fig:Fig7}a. On the other hand, for
$p/\varepsilon = 1/3$, we are getting $dV/dT$, as in the Fig.
\ref{fig:Fig7}b. 

We would like to close with the following remark. Our result
presented in Fig. \ref{fig:Fig7} shows not one but two "soft points" 
or "critical points" of EoS (so far we cannot distinguish here
between these two possibilities), located at different energies: one
at $\sqrt{s} \sim 6$ GeV (obtained from data on $V$ at different
energies \cite{PRL} and from data on $\Phi$ for different
centralities \cite{NA49}) and another one at $\sqrt{s} \sim 100$ GeV
(obtained from data on fluctuations measured by quantity $F_T$ for
different centralities \cite{PHENIX}). We cannot so far offer
explanation of this result. Interpreting it, however, in the original
spirit of looking for the possible signals of QGP (but keeping in
mind our reservation concerning this point expressed above) we could
say that it looks like, with increasing energy, first occurs a kind
of QGP composed mostly of dressed quarks, which at higher energies is
followed by a true QGP containing also liberated gluons
\footnote{Notice that data from \cite{PRL} indicate that volume of
interaction grow much slower than linearly with energy, i.e., the
energy densities obtained at different energies increase
substantially: $\varepsilon \sim 0.1s^{0.41}$.}. What we see here
most probably indicates some additional change in energy dependence
of $V$ presented in Fig. \ref{fig:Fig6}a, which would take place at
$\sqrt{s} \sim 100$ GeV. If true it would indicate that $V$ is not
increasing with energy since that point, on the contrary, it probably
decreases a little (what seems to be confirmed by the first data from
RHIC at $200$ GeV \cite{RHIC}). One should also keep in mind, before
further speculations, that the assumed here access to the EoS by use
of fluctuations in hadronic production neglects altogether the
possible evolution processes between the QGP and state of freely
streaming hadrons observed experimentally. This remark applies also
to all works of this kind, like \cite{GGM}. It means therefore that
final conclusions could only be drawn when other approaches, as for
example that discussed in \cite{PERCOL}, are also confronted with all
available data, as has been done here.\\

\appendix
\section{}
We shall, for completeness, derive here formula (\ref{eq:R}). We
shall consider in what follows our hadronizing system as statistical
system in equilibrium without specifying its details. Thermodynamical
potential of such system is
\begin{equation}
\Omega\, =\, E\, +\, pV\, -\, TS\, =\, \mu N. \label{eq:Phi}
\end{equation}
It is then connected with chemical potential $\mu$ with $N$ denoting
number of particles in the system under consideration. For system in
equilibrium but with varying number of particles $N$ is given by
condition $d\Omega/dN=0$ or, equivalently, $\mu =0$. For conditions of
statistical equilibrium we have then (here $\varepsilon$ and $s$ are
the corresponding densities of energy and entropy)
\begin{equation}
\varepsilon\, +\, p\, -\, Ts\, =\, 0 . \label{eq:eps}
\end{equation}
Using (\ref{eq:eps}) together with thermodynamic identity
\begin{equation}
dE\, =\, -pdV\, +\, TS \label{eq:dE}
\end{equation}
we get
\begin{equation}
d\varepsilon\, +\, dp\, =\, Tds\, +\, sdT . \label{eq:tds}
\end{equation}
Denoting $\alpha = c_0^2=dp/d\varepsilon = p/\varepsilon$ we obtain
from (\ref{eq:eps}) and (\ref{eq:tds}) 
\begin{equation}
\frac{ds}{s}\, =\frac{1}{\alpha}\frac{dT}{T}\qquad {\rm and}\qquad 
d\varepsilon\, =\, \frac{s}{\alpha}dT . \label{eq:dsde}
\end{equation}
Because from (\ref{eq:eps}) we get $\varepsilon = Ts-p =
T\alpha(d\varepsilon/dT) - \alpha \varepsilon$, then
\begin{equation}
\frac{d\varepsilon}{\varepsilon}\, =\,
\frac{1+\alpha}{\alpha}\frac{dT}{T} . \label{eq:eovere}
\end{equation}
From equations (\ref{eq:dsde}) and (\ref{eq:eovere}) one gets that
ratio of fluctuations of {\it densities} is given by
\begin{equation}
R\, =\, \frac{(ds/s)^2}{d\varepsilon/\varepsilon)^2}\, =
\frac{1}{(1 + \alpha )^2} . \label{eq:Rdensities}
\end{equation}
Because $S=sV$ and $E=\varepsilon V$ and, respectively, $dS = ds V +
s dV$ and $dE = Vd\varepsilon + \varepsilon dV$, then
\begin{equation}
\frac{dS}{s} = \frac{1}{\alpha}\frac{dT}{T}\left(1 +
\alpha\frac{T}{V}\frac{dV}{dT}\right)\quad {\rm and}\quad
\frac{dE}{E} = \frac{1+\alpha}{\alpha}\frac{dT}{T}\left(1 +
\frac{\alpha}{1+\alpha}\frac{T}{V} \frac{dV}{dT}\right), \label{eq:SE}
\end{equation}
and the corresponding ratio of fluctuations of entropy and energy is
equal 
\begin{equation}
R\, =\, \frac{(dS/S)^2}{(dE/E)^2}\, =\,  \left(1\, +\,
\frac{\alpha}{1+\alpha \phi}\right)^{-2}\quad {\rm where}\quad
\phi\, =\, \frac{T}{V}\frac{dV}{dT}\, =\, \frac{d\ln V}{d \ln T} .
\label{eq:result}
\end{equation}
For $dV=0$ one gets result obtained already in \cite{GGM}. However,
in general $\phi \neq 0$ and fluctuations of the interaction volume
are also important. Eq. (\ref{eq:result}) can be also rewritten as
\begin{equation}
\frac{1}{\alpha}\, =\, -\phi\, +\, \frac{\sqrt{R}}{1-\sqrt{R}} ,
\label{eq:RR} 
\end{equation}
which connects EoS (represented by $\alpha$) with changes in the
reaction volume and with fluctuations  described by the parameter
$R$. 

\section{}
We shall derive here eq.(\ref{eq:FRF}). Let us first notice that $F_T$
defined in (\ref{eq:FT}) is related to other measure of fluctuations,
the so called $\Phi$-measure introduced in \cite{PHI} and used in
\cite{NA49,CERES},    
\begin{equation}
\Phi = \sqrt{\frac{\left\langle Z^2\right\rangle}{\langle N\rangle}} -
\sqrt{\left\langle z^2\right\rangle} . \label{eq:PHI}
\end{equation}
For a measured quantity $x$ (here identified with transverse momentum
of produced particles, $x_i = p_{Ti}$) one has: 
\begin{eqnarray}
z &=& x - \langle x\rangle,\quad \left\langle z^2\right\rangle =
Var(x) = \left\langle x^2\right\rangle - \langle x\rangle^2\nonumber\\
Z &=& \sum_{i=1}^N z_i = \sum_{i=1}^N x_i - N\langle x\rangle, \label{eq:def}\\
\left\langle Z^2\right\rangle &=& \left\langle
\left(\sum_{i=1}^N x_i\right)^2\right\rangle - 2\langle
x\rangle\left\langle N\sum_{i=1}^Nx_i\right\rangle + \left\langle
N^2\right\rangle\left\langle x\right\rangle^2.\nonumber
\end{eqnarray}
To get this result one uses following reasoning. If variables $x$ and
$N$ are described by distributions characterized by the respective 
generating functions $f(t)$ and $h(t)$ then 
and variable $N$ by generating function $h(t)$ then variable
$\xi=\sum_{i=1}^N x_i$ is described by generating function 
$G(t) = h[f(t)]$. It means therefore that  
\begin{equation}
\langle \xi\rangle = G'(1)=\langle N\rangle \langle x\rangle \label{eq:one}
\end{equation}
and
\begin{equation}
\langle \xi^2\rangle = G''(1) + G'(1) = \langle N^2\rangle\langle
x\rangle^2 + \langle N\rangle Var(x) \label{eq:two}
\end{equation}
leading to 
\begin{equation}
Var(\xi) = \langle N\rangle Var(x) + \langle x\rangle^2 Var(N) . \label{eq:three}
\end{equation}
To characterize correlations between variables $N$ and $\xi$ one has
to introduce a correlation coefficient $\rho \in [-1,1]$, which will
be, in what follows, our free parameter (our minimal input). 
One can therefore write:
\begin{equation}
\left\langle N\sum_{i=1}^Nx_i\right\rangle = \langle N\rangle
\left\langle \sum_{i=1}^Nx_i\right\rangle + \rho \sqrt{Var(N)
Var\left(\sum_{i=1}^Nx_i\right)} . \label{eq:rho}
\end{equation}
The last term in (\ref{eq:rho}) can be written as
\begin{equation}
\sqrt{Var(N)Var\left(\sum_{i=1}^Nx_i\right)} =
\sqrt{\langle N\rangle Var(N)Var(x) + \langle x\rangle ^2
Var^2(N)}. \label{eq:uno} 
\end{equation}
Substituting now (\ref{eq:rho}) to (\ref{eq:def}) and making use of
eq. (\ref{eq:one})-(\ref{eq:three}) one gets
\begin{equation}
\langle Z^2\rangle = \langle N\rangle Var(x) + 
+ 2\langle x\rangle ^2 Var(N)\left[1 - \rho\sqrt{
   \frac{\langle N\rangle}{Var(N)}\frac{Var(x)}{\langle x\rangle ^2}
+ 1 }\right] . \label{eq:Z}
\end{equation}
Substituting this to eq. (\ref{eq:PHI}) and making use of the
relation between $F_T$ and $\Phi$ derived in \cite{FOOTB}, namely
that 
\begin{equation}
F_T = \frac{\Phi}{\sqrt{\left\langle z^2\right\rangle}} ,
\label{eq:FTPHI}
\end{equation}
one gets 
\begin{equation}
F_T = -1 + \sqrt{1+
 2\frac{\langle x\rangle^2}{Var(x)}\frac{Var(N)}{\langle N\rangle}
 \left[1 - \rho\sqrt{
   \frac{\langle N\rangle}{Var(N)}\frac{Var(x)}{\langle x\rangle ^2}
   + 1 }\right] }, \label{eq:Ft}
\end{equation}
which is eq.(\ref{eq:FRF}) we were looking for.

\section{}
Let us start with change in the energy $E$ of the system, which can
be written as ($C_V$ is corresponding heat capacity)
\begin{equation}
\Delta E\, =\, \left(\frac{\partial E}{\partial V}\right)_T\Delta V
+ \left(\frac{\partial E}{\partial T}\right)_V\Delta T =
\left[T\left(\frac{\partial p}{\partial T}\right)_V - p\right]\Delta
V + C_V\Delta T . \label{eq:DE}
\end{equation}
Squaring it and averaging while remembering that \cite{Landau}
fluctuations of the volume and temperature are given by, respectively,
\begin{equation}
\overline{(\Delta V)^2} = - T\left(\frac{\partial V}{\partial p}\right)_T
\quad{\rm and}\quad \overline{(\Delta T)^2} = \frac{T^2}{C_V} \label{eq:VT}
\end{equation}
and are statistically independent, i.e., $\overline{\Delta T\delta V} =0$,
one gets
\begin{equation}
Var(E)\, =\, \overline{\left(\Delta E^2\right)}\, =\, -\left[T\left(\frac{\partial
p}{\partial T}\right)_V - p\right]^2\, T\left(\frac{\partial
V}{\partial p}\right)_T + C_VT^2 . \label{eq:VarE}
\end{equation}
Similarly, fluctuations of number of particles can be described by formula
\begin{equation}
Var(N)\, =\, - \frac{T\langle N\rangle^2}{V^2}\left(\frac{\partial
V}{\partial p}\right)_T . \label{eq:VarN}
\end{equation}
From (\ref{eq:VarE}) and (\ref{eq:VarN}) one gets 
\begin{equation}
R\, =\, \frac{ \frac{Var(N)}{\langle N\rangle^2} }
             { \frac{Var(E)}{\langle E\rangle^2} }\, =\, 
             \frac{1}{\left[\frac{T}{\varepsilon}
             \left(\frac{\partial p}{\partial T}\right)_V - 
                   \frac{p}{\varepsilon} \right]^2
              - \frac{C_VT}{\varepsilon^2}
             \left(\frac{\partial p}{\partial V}\right)_T} . \label{eq:RAP}
\end{equation}
Because
\begin{equation}
T\left(\frac{\partial p}{\partial T}\right)_V\, =\,
\left(\frac{\partial E}{\partial V}\right)_T + p \label{eq:TEP}
\end{equation}
then (from now on all derivatives are for $T=const$)
\begin{eqnarray}
R\, &=&\, \frac{\varepsilon^2}{\left[
           \left( \frac{\partial E}{\partial V}\right)^2 - 
       C_VT\left( \frac{\partial p}{\partial V}  \right)
                              \right]}\, =\, 
           \frac{\varepsilon^2
              \left(\frac{\partial V}{\partial E}\right)^2}
                {1 - C_V T \frac{\partial p}{\partial V}
                           \frac{\partial V}{\partial E}
                           \frac{\partial V}{\partial E}}\, =\nonumber\\
    &=&\, \frac{\varepsilon^2 \left(\frac{\partial V}{\partial E}\right)^2}
           {1-C_VT\frac{\partial p}{\partial E}
                  \frac{\partial V}{\partial E}} .     \label{eq:Rap}
\end{eqnarray}
Because of EoS one has that 
\begin{equation}
p\, =\, \alpha \varepsilon\qquad {\rm and}\qquad \frac{\partial
p}{\partial E}\, =\, \frac{\partial p}{\partial V}\frac{\partial
V}{\partial E}\, =\, \frac{\alpha}{V}\, >\, 0 , \label{eq:EOS}
\end{equation}
what immediately leads to eq.(\ref{eq:RRRR}):
\begin{equation}
R = \frac{\varepsilon^2\left(\frac{\partial V}{\partial E}\right)^2}
    {1 - C_VT\frac{\alpha}{V}\frac{\partial V}{\partial E}}.
\label{eq:RRRRA}
\end{equation}

Partial support of the Polish State Committee for Scientific Research
(KBN) (grant 2P03B04123 (MR and ZW) and grants
621/E-78/SPUB/CERN/P-03/DZ4/99 and 3P03B05724 (GW)) is acknowledged.


\begin{thebibliography}{99}

\bibitem{QM} H.Gutbrod, J.Aichelin and K.Werner (Edts), 
             {\sl Nucl. Phys.} {\bf A715} (2003). 
                                                             
\bibitem{GGM} M.Ga\'zdzicki, M.I.Gorenstein and S.Mr\'owczy\'nski,
              {\it Fluctuations and Deconfinement Phase Transition in
              Nucleus - Nucleus Collisions}, hep-ph/0304052. 

\bibitem{SSS} L.Stodolsky, {\sl Phys. Rev. Lett.}
              {\bf 75} (1995) 1044; St.Mr\'owczy\'nski, 
              {\sl Phys. Lett.} {\bf B430} (1998) 9; 
              M.Stephanov, K.Rajagopal and E.Shuryak, {\sl Phys. Rev.
              Lett.} {\bf 81} (1998) 4816 and  {\sl Phys. Rev.} 
              {\bf D60} (1999) 114028.
                                                   
\bibitem{THM} T.A.Trainor, {\it Event-by-Event analysis and the 
              Central Limit Theorem}, hep-ph/0001148; H.Heiselberg, 
              {\sl Phys. Rep.} {\bf 351} (2001) 161; B.M\"uller,
                {\sl Nucl. Phys.} {\bf A702} (2002) 281c.
                
\bibitem{LML} L.McLerran, {\it RHIC Physics: The Quark Gluon Plasma
              and the Color Glass Condensate: 4 Lectures},
              hep-ph/0311028. 

\bibitem{GG} M.Ga\'zdzicki, {\it Energy Scan Program at the CERN SPS 
             and an Observation of the Deconfinement Phase Transition 
             in Nucleus-Nucleus Collisions}, presented at $7$th Int. 
             Conf. on Strange Quarks in Matter, SQM 2003, March 2003, 
             Atlantic Beach, USA; hep-ph/0305176 and {\sl Acta Phys. 
             Polon} {\bf B34} (2003) 5771;
             M.I.Gorenstein, {\it Signals of Deconfinement Transition
             in Nucleus-Nucleus Collisions}, hep-ph/0310269.


\bibitem{PERCOL} E.G.Ferreiro, F.del Moral and C.Pajares, 
                 {\it Transverse momentum fluctuations and
                 percolation of strings}, hep-ph/0303137.

\bibitem{PHENIX} J.Nystrand (PHENIX Coll.),
                 {\sl Nucl. Phys.} {\bf A715} (2003) 603c.

\bibitem{NA49} C.Blume et al (NA49 Coll.) {\sl Nucl.Phys.} 
               {\bf A715} (2003) 55c;
               T.Anticic et al., (NA49 Coll.), {\it Transverse 
               momentum fluctuations in nuclear
               collisions at $158$ AGeV}, hep-ex/0311009.

\bibitem{HS} C.M.Hung and E.V.Shuryak, {\sl Phys. Rev. Lett.} {\bf
             75} (1995) 4003.

\bibitem{LAT} C.R.Alton et al., {\it The Equation od State for Two
              Flavor QCD at Non-zero Chemical Potential},
              hep-lat/0305007. See also Z.Fodor, {sl Nucl. Phys..}
              {\bf A715} (2003) 319c.

\bibitem{Spinod} J.Randrup, {\it Spinodial decomposition during the
                 hadronization stage at RHIC?}, hep-ph/030827.

\bibitem{CERES} D.Adamova et al., (CERES Coll.) {\sl Nucl.Phys.}
                {\bf A727} (2003) 97.

\bibitem{PHI} M.Ga\'zdzicki and S.Mr\'owczy\'nski, {\sl Z. Phys.}
              {\bf C54} (1992) 127.

\bibitem{PRL} D.Adamova et al. (CERES Coll.), {\sl Phys. Rev. Lett.}
              {\bf 90} (2003) 022301. 

\bibitem{FOOTB} K.Adcox et al. (PHENIX Collab.), {\sl Phys. Rev.} 
                {\bf C66} (2002) 024901.

\bibitem{PhiUWW} O.V.Utyuzh, G.Wilk and Z.W\l odarczyk,
                 {\sl Phys. Rev.} {\bf C64} (2001) 027901.
                
\bibitem{Broad} M.M.Aggarwal et al (WA98 Coll.), {\sl Phys. Rev.} 
                {\bf C65} (2002) 054912. 

\bibitem{NB} See, for example, P.Carruthers and C.C.Shih, {\sl Int.
             J. Mod. Phys.} {\bf A2} (1986) 1447 or C.Geich-Gimbel,
             {\sl Int. J. Mod. Phys.} {\bf A4} (1989) 1527.

\bibitem{WWq} G.Wilk and Z.W\l odarczyk, {\sl Phys. Rev. Lett.} {\bf 
              84} (2000) 2770. 

\bibitem{WWqq} G.Wilk and Z.W\l odarczyk,
               {\sl Chaos, Solitons and Fractals} {\bf 13} (2002) 581
               and {\sl Physica} {\bf A305} (2002) 227.

\bibitem{INELK} F.S.Navarra, O.V.Utyuzh, G.Wilk and Z.W\l odarczyk,
                {\sl Phys. Rev.} {\bf D67} (2003) 114002.

\bibitem{Huang} Cf., for example, K.Huang, {\it Statistical
                Mechanics}, John Wiley and Sons, New York (1987).

\bibitem{Landau} L.D.Landau, E.M.Lifshitz, {\it Course of Theoretical
                 Physics: Statistical Physics}, Pergamon Press, New
                 York 1958. 

\bibitem{RHIC} M.L\'opez Noriega (STAR Coll.), {\sl Nucl. Phys.} 
               {\bf A715} (2003) 623c.

\end{thebibliography}
\end{document}